%% file: regnann_grimaldi.tex
\documentclass{bioinfo}
\usepackage{graphicx,url}
\copyrightyear{2010}
\pubyear{2010}
\history{}
\editor{}

\newcommand{\ignore}[1]{}

\begin{document}

\firstpage{1}
\title[RegnANN: Variability in Network Inference Algorithms]{Reverse Engineering Gene Networks with ANN:\\Variability in Network Inference Algorithms}
\author[Grimaldi \textit{et al.}]{Marco Grimaldi\,$^{\rm a}$\footnote{to whom correspondence should be addressed.} , Giuseppe Jurman\,$^{\rm a}$, Roberto Visintainer$^{\rm a,b}$}
\address{
$^{\rm a}$ Fondazione Bruno Kessler, via Sommarive 18, I-38123 Povo (Trento), Italy,\\
$^{\rm b}$ DISI, University of Trento, via Sommarive 14, I-38123 Povo (Trento), Italy}
\maketitle

\begin{abstract}
\section{Motivation:}
Reconstructing the topology of a gene regulatory network is one of the key tasks in systems biology. Despite of the wide variety of proposed methods, very little work has been dedicated to the assessment of their stability properties.
Here we present a methodical comparison of the performance of a novel method (RegnANN) for gene network inference based on multilayer perceptrons with three reference algorithms (ARACNE, CLR, KELLER), focussing our analysis on the prediction variability induced by both the network intrinsic structure and the available data.

\section{Results:}
The extensive evaluation on both synthetic data and a selection of gene modules of \emph{Escherichia coli} indicates that all the algorithms suffer of instability and variability issues with regards to the  reconstruction of the topology of the network. 
This instability makes objectively very hard the task of establishing which method performs best. Nevertheless, RegnANN shows MCC scores that compare very favorably with all the other inference methods tested.

\section{Availability:}
The software for the RegnANN inference algorithm is distributed under GPL3 and it is available at the corresponding author home page \url{http://mpba.fbk.eu/grimaldi/regnann-supmat}, together with the Supplementary Material.

\section{Contact:} \href{grimaldi@fbk.eu}{grimaldi@fbk.eu} (Marco Grimaldi)
\end{abstract}

\section{Introduction}
\label{sec:introduction}
%The research trend in systems biology is experiencing a shift
%from the analysis of the single gene activity to the understanding of
%the complex network of their mutual connections (called Gene
%Regulatory Network, GRN for short). 
Since the first examples dating
back to early seventies (\cite{glass73logical}), the
challenge of reconstructing the links among genes in a regulatory network starting from their
expression signals has been tackled by several laboratories worldwide. 
These initial efforts have originated a number of related publications which has been
exponentially growing in the last few years.

The inference methods generally employed are of very different nature, ranging from
deterministic, e.g.: systems of differential equations (\cite{bansal07how})
and Groebner bases (\cite{dimitrova07groebner}),
to stochastic approaches, e.g.:  Boolean (\cite{kauffman93origins}) or Bayesian
(\cite{friedman00using}) algorithms. Such approaches may also start from different types
of gene expression data: time-course or steady states. Furthemore, also the detail and the complexity of the considered network can 
vary as the links may carry information about the direction of the relation (directed graph) and a weight may be associated to the strenght of each link (weighted directed graph) \cite{markowetz07inferring, karlebach08modelling}.
Generally, the reconstruction accuracy is far from being optimal in many
situations with the presence of several pitfalls, related to both the methods and the
available data (\cite{he09reverse}).  Citing \cite{baralla09inferring}, "Inferring gene networks is a daunting
task", not only in terms of devising an effective algorithm, but also in terms
of quantitatively interpreting the obtained results.  Only recently 
efforts have been carried out towards an
objective comparison of network inference methods also highlighting
occurring limitations (\cite{krishnan07indeterminacy,
altay10revealing, marbach10revealing}).

This work compares four network reverse engineering methods, first settling in a
controlled situation with synthetic data and then focusing on a biological setup by analysing 
transcriptional subnetworks of \emph{Escherichia coli}. 
In order to simplify our comparative evaluation, we will only consider the underlying topology, thus neglecting both weight and direction of the links among the genes. 
In doing so, we confine the analysis of the reconstructed network in terms of the binary existence or not-existence of an edge.
The general performance of the network inference task is evaluated in terms of Matthews Correlation Coefficient
(MCC, \cite{matthews75comparison} -- see Sup. Mat. for details). MCC is becoming the 
measure of choice in many application fields of machine learning and bioinformatics: it is one of the best methods for summarizing into a single
value the confusion matrix of a binary classification task. Recently it has also been used for comparing network topologies (\cite{stokic09fast}).
 
In this paper we introduce a novel inference method called Reverse
Engineering Gene Networks with Artificial Neural Networks (RegnANN). This approach is based on
an ensemble of multilayer perceptrons trained using steady state data.  Its perfomance
is compared with those of top-scoring methods such as KELLER
(\cite{song09keller}), ARACNE (\cite{margolin06aracne}) and CLR
(\cite{faith07large}) while assessing possible
sources of instability. To improve the general efficiency of RegnANN we
implement the algorithm using GPGPU (\cite{lahabar08high})%, scanzio10parallel}).

The extensive evaluation on both synthetic and biological data indicates that the algorithms tested suffer of instability and 
variability issues with regards to the reconstruction of the network topology. The instability makes objectively very hard the 
task of establishing which method performs best. Nevertheless, RegnANN shows MCC scores that compare very favorably with all 
the other inference methods tested.

\begin{methods}
\section{Methods}
\label{sec:methods}
\subsection{RegnANN: network inference using ANN}
\label{ssec:inference}
To infer gene regulatory networks we adopt an ensemble of feed-forward multilayer perceptrons (\cite{bishop95neural}) trained using the back-propagation algorithm. 
Each member of the ensemble is essentially a multi-variable regressor (one to many) trained using an input expression matrix to learn the relationships (correlations) among a target gene and all the other genes in the network.
We proceed in determining the interactions among genes separately and then we join the information to form the overall network. 
From each row of the gene expression matrix\footnote{In this work we consider gene expression matrices of dimension $M \times N$: $N$ genes whose expression levels are recorded $M$ times.} we build a set of input and output patterns used to train a selected multilayer perceptron. 
Each input pattern corresponds to the expression value for the selected gene of interest. 
The output pattern is the row-vector of expression values for all the other genes for the given row in the gene expression matrix (Figure \ref{fig:regnann}).
By cycling through all the rows in the matrix, each regressor in the ensemble is trained to learn the correlations among one gene and all the others. 
Repeating the same procedure for all the columns in the expression matrix, the ensemble of multi-variable regressors is trained to learn the correlations among all the genes.

The procedure of determining separately the interactions among genes is very similar to the one presented in \cite{song09keller}, where the authors propose to estimate the \emph{neighborhood} of each gene (the correlations among one gene and all the others) independently and then joining these neighborhoods to form the overall network, thus reducing the problem to \emph{a set of identical atomic optimizations} (Section \ref{ssec:othermethods}).
 
Here we build $N$ -- one for each of the $N$ genes in the network -- multilayer perceptrons with one input node, one layer of hidden nodes and one layer of $N-1$ output nodes. 
The input node takes the expression value of the selected gene rescaled in $\left[-1, 1\right]$. 
The number of hidden nodes is set empirically to the square root of the number of inputs by the number of outputs, resulting in $\sqrt{N-1}$. 
The activation function is the hyperbolic tangent, which provides output values in the range $\left[-1, 1\right]$, thus making the output values interpretable in terms of positive correlation ($+1$), anti-correlation ($-1$)  and not-correlated ($0$). 
The other parameters used to learn each multi-layer perceptron are as follows: learning rate equal to $0.01$; momentum equal to $0.1$, learning epochs equal to $10000$; bias equal to $0$\footnote{These values are evaluated empirically during preliminary tests on synthetic data.}.
 
Finally, the topology of gene regulatory networks is obtained by applying a second procedure. The correlation of each gene with all the others is extracted by passing a purposely made test pattern to the regressor: considering separately each multilayer perceptron in the ensemble, a value of $1$ is passed to its input neuron, consequently recording its output values. In this way, the correlation between the corresponding gene with all the others is obtained as a vector of values in $\left[-1, 1\right]$. By cycling through all the members of the regression system, we obtain the adjacency matrix of the sought gene network. It is important to note that this procedure does not allow discovering of gene self correlation (regulation) patterns, but only correlation patterns among different genes. Moreover, the algorithm here proposed cannot estimate future values, because it is not a \emph{predictor}, as in the case of GRNN (\cite{specht93general}: instead it \emph{models} static correlations between genes. As in \cite{song09keller}, it is possible to extend the regression system to take into account dynamic rewiring of the topology, but this is beyond the scope of the present work.
 
To improve the general efficiency of the algorithm and thus allow a systematic comparison of its performance with the other gene network reverse engineering methods tested (Subsection \ref{ssec:othermethods}), we implemented the ANN based regression system using the GPGPU programming paradigm (\cite{lahabar08high, scanzio10parallel}).

%%%%%%%%%%%%%%%%%%%%%%%%%%%%%%%%%%%%%%%%%%%
\begin{figure}[!t]
\centerline{\includegraphics[scale=0.40]{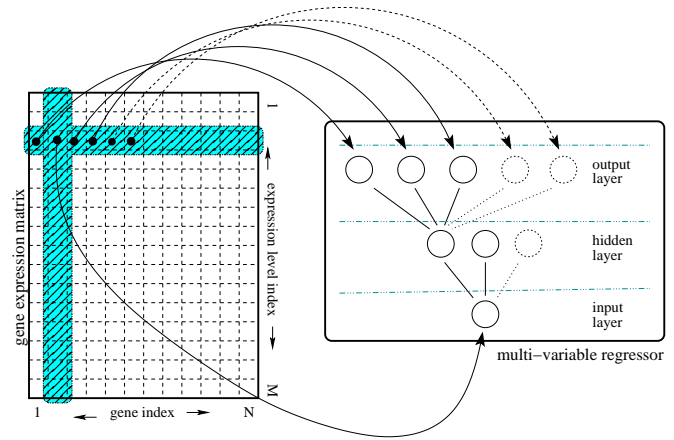}}
\caption{The \emph{ad hoc} procedure proposed to build the training input/output patterns starting from a gene expression matrix. Each input pattern corresponds to the expression value for the selected gene of interest. The corresponding output pattern is the vector of expression values for all the other genes for the given row in the gene expression matrix.}
\label{fig:regnann}
\end{figure}
%%%%%%%%%%%%%%%%%%%%%%%%%%%%%%%%%%%%%%%%%%%

\subsection{Alternative inference methods} 
\label{ssec:othermethods}
As reference methods we select three alternative algorithms widely used in literature: ARACNE, CLR and KELLER.
\paragraph{KELLER:} 
it is  a kernel-reweighted logistic regression method (\cite{song09keller}) introduced for reverse engineering the dynamic interactions between genes based on the time series of their expression values. It estimates the neighborhood of each gene separately and then joins the neighborhoods to form the overall network. The approach aims at reducing the network inference problem to a set of identical atomic optimizations. KELLER makes use of the $l_{1}$-regularized logistic regression algorithm and operates modeling the distribution of interactions between genes as a binary pair-wise Markov Random Field. The method has been applied to reverse engineer genome-wide interactions taking place during the life cycle of \emph{Drosophila melanogaster}. 
Although KELLER has been developed to uncover dynamic rewiring of gene transcription networks (e.g.: dynamic changes in their topology), here we consider constant network topology for a given gene expression matrix. In this work we make use of the reference implementation of the algorithm provided in \cite{song09keller}.
\paragraph{ARACNE:} 
it is a general method able to address a wide range of network deconvolution problems -- from transcriptional (\cite{margolin06aracne}) to metabolic networks (\cite{nemenman07reconstruction}) -- that was originally designed to scale up to the complexity of regulatory networks in mammalian cells. The method makes use of an information theoretic approach to eliminate the majority of indirect interactions inferred by co-expression methods. ARACNE removes the vast majority of indirect candidate interactions using a well-known information theoretic property: the data processing inequality (\cite{cover91elements}).
In this work we use the reference implementation of the algorithm provided in \cite{meyer08minet} with default value for the data processing inequality tolerance parameter.
\paragraph{CLR:}
it is an extension of the relevance networks class of algorithms (\cite{faith07large}), which predicts regulations between transcription factors and genes making use of the mutual information score. CLR proposes an adaptive background correction step that is added to the estimation of mutual information. For each gene, the statistical likelihood of the mutual information score is computed within its network context. Then, for each transcription factorÐ-target gene pair, the mutual information score is compared to the context likelihood of both the transcription factor and the target gene, and turned into a z-score. We adopt the reference implementation of the algorithm provided in \cite{meyer08minet}.

\subsection{Experimental protocol}
\label{ssec:protocol}
We are interested in comparing the performance of the selected reverse engineering methods in inferring the underlying topology of regulatory networks. As proposed in \cite{song09keller}, we focus on the estimation of the interaction structures between genes, rather than the strength of these interactions. The inferred adjacency matrix is symmetric and discretized with values in $\{0,1\}$ by thresholding. 

The binarization of the inferred network obtained with RegnANN is achieved using by using a threshold value of $0.5$. In the case of KELLER, the reference implementation (\cite{song09keller}) returns a symmetric and discrete (with values in $\{ 0,1 \}$) adjacency matrix -- binarization is obtained by rounding values bigger than $10^{-3}$ to $1$. Results obtained with ARACNE are discretized as in the case of KELLER. Usually, the cutoff value for the mutual information is estimated for each data-set separately using a significance measure  (e.g.: the F-score (\cite{altay10revealing})) or building a Precision-Recall curve and selecting the desired threshold value (\cite{margolin06aracne}). Here, the threshold value is kept constant to avoid the introduction of a selection bias in the outcome of the ARACNE algorithm. The same procedure is applied to CLR (threshold value of $10^{-3}$).

The accuracy (in terms of MCC) of the inference methods is firstly evaluated on synthetic data (Section \ref{sec:data}) by varying the topology of the network, its size, the amount of data available, the method adopted to synthesize the data and the method adopted to normalize the data prior to network inference -- see Supplementary Material for details. 
Methodically, we vary one parameter at a time and then measure the performance of the systems as the mean of $10$ randomly initialized runs. For each run, the network topology is randomly generated with the desired number of genes ($N$), the expression profiles -- the data -- are (randomly) generated the required number of times ($M$), the selected normalization method is applied and the MCC values for the applied reverse engineering method recorded. The error of the measurement is expressed as twice the standard deviation of the $10$ independent runs.

Finally, the performance of the four network inference algorithms is tested on $7$ selected gene network modules of \emph{Escherichia coli} (\cite{alvarez09modular}). While ARACNE, CLR and KELLER are deterministic algorithms\footnote{Given a particular input, the algorithm will always produce the same output, always passing through the same sequence of states.}, RegnANN may produce different results depending on the random initialization of the weights in the ensemble of multi-layer perceptrons. Thus, in order to smooth out possible local minima, we adopted a majority voting schema: for each network module, the RegnANN algorithm is applied $10$ times and the inferred adjacency matrices accumulated. The final topology is obtained selecting those links that appeared with a frequency higher than $7$ (out of $10$). The entire procedure is repeated $10$ times and the final prediction is estimated as the mean and the associated error as twice the standard deviation of the $10$ independent runs.
\end{methods}

\section{Data}
\label{sec:data}
\paragraph{Synthetic data:} we benchmark the reverse engineering algorithms here considered using both synthetic and biological data. Synthetic data are obtained considering two different network topologies: Barabasi-Albert (\cite{barabasi99emergence}) and Erd\"os-R\'enyi (\cite{erdos59random}). Furthermore, we apply two different gene expression synthesis methods: the first one considers only linear correlation among selected genes (SLC), the second one is based on a gene network/expression simulator recently proposed to assess reverse engineering algorithms (GES, \cite{dicamillo09gene}). See Supplementary Material for full details.

\paragraph{Escherichia coli data:} the task for the biological experiments is the inference of a few transcriptional subnetworks of the model organism \emph{Escherichia coli} starting from a set of steady state gene expression data.
The data are obtained from different sources and they consist of three
different elements, namely the whole \emph{Escherichia coli} transcriptional
network, the set of the transcriptional subnetworks and the gene
expression profiles to infer the subnetworks from.  The \emph{Escherichia coli}
transcriptional network is extracted from the
RegulonDB\footnote{\url{http://regulondb.ccg.unam.mx/}} database,
version 6.4 (2010) and it consists of 3557 experimentally confirmed
regulations between 1442 genes, amongst which 172 transcription
factors.  The 117 subnetworks are defined in \cite{marr10patterns}: in
our experiments we use 7 of these subnetworks, including a number of
genes ranging from 7 to 104.  The expression data have been originally
used in \cite{faith07large} and consist of 445 \emph{Escherichia coli}
Affymetrix Antisense2 microarray expression profiles for 4345 genes,
collected under different experimental conditions such as PH changes,
growth phases, antibiotics, heat shock, varying oxygen concentrations
and numerous genetic perturbations.  MAS5 preprocessing is chosen among the available
options (MAS5, RMA, gcRMA, DChip).

\section{Results}
\label{ssec:results}
Due to space constraints, hereafter we present a selection of the outcomes of the experimental evaluation with emphasis on the reconstruction variability; for previous usage of MCC in network theory and applications see \cite{stokic09fast, supper07reconstructing}.  

\subsubsection*{Synthetic data:}
\label{ssec:synth}
\begin{figure}[!t]
\centerline{\includegraphics[scale=0.42]{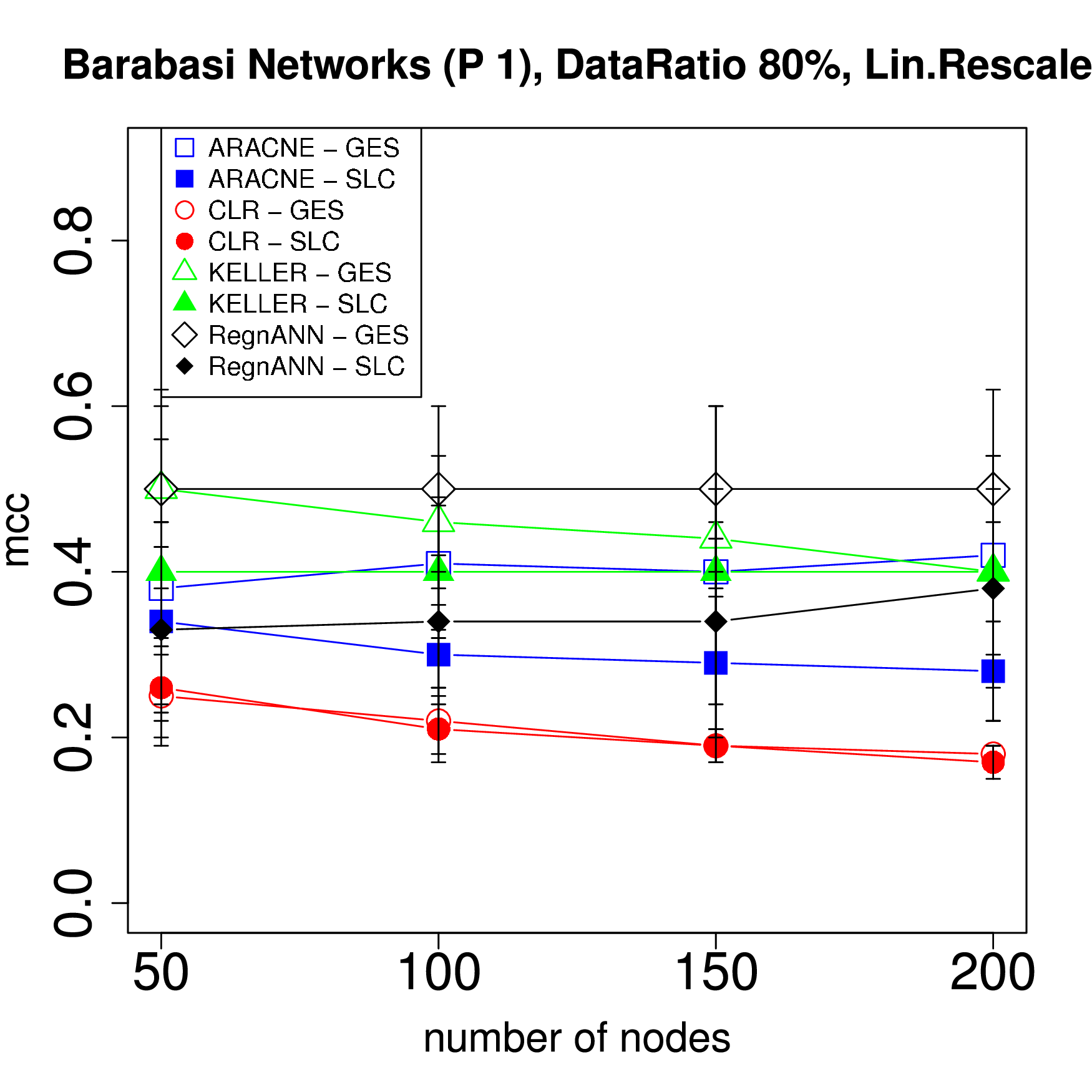}}
\caption{MCC scores of the different network inference algorithms for synthetic Barabasi networks (scale free, exponent $P=1$), varying the number of nodes and keeping constant the data ratio: the number of expression profiles to the number of nodes ($80\%$). Both methods (GES and SLC) for data synthesis are considered. Expression values are linearly rescaled in $[-1,1]$.}
\label{fig:Barabasi_mcc_vs_nodes}
\end{figure}

Figure \ref{fig:Barabasi_mcc_vs_nodes} illustrates the MCC scores obtained with ARACNE, CLR, KELLER and RegnANN for synthetic Barabasi networks (scale free, exponent $P=1$), varying the number of nodes. In order to provide similar amount of information to the inference algorithms while varying the size of the network, we kept constant the \emph{data ratio}: the number of expression profiles to number of nodes ($80\%$) -- e.g.: $50$ nodes, 40 different expression profiles; 200 nodes 160 different expression profiles. Expression values are linearly rescaled in $[-1,1]$.
%
%%%%%%%%%%%%%%%%%%%%%%%%%%%%%%%%%%%%%%%%%%%
%%%%%%%%%%%%%%%%%%%%%%%%%%%%%%%%%%%%%%%%%%%
%
Figure \ref{fig:Barabasi_mcc_vs_nodes} indicates that the MCC scores on Barabasi networks depend on both the inference algorithm and the data synthesis methods, while the size of the network (number of nodes considered) has a somewhat smaller impact on the performance. RegnANN-GES scores $0.5 \pm 0.1$ on a network of $200$ nodes, while RegnANN-SLC scores $0.34 \pm 0.08$ on a similarly sized network. KELLER scores $0.4 \pm 0.1$ irrespective of the data synthesis method applied on the 200 nodes network. On the same sized network, ARACNE-GES scores $0.42 \pm 0.04$ while ARACNE-SLC scores $0.28 \pm 0.06$. Finally, CLR shows the worst performance of the four algorithms tested, irrespective of the network size and the data synthesis adopted, e.g.: $0.17 \pm 0.02$ (GES) for a network of $200$ nodes -- $0.18 \pm 0.01$, in the case of SLC.

\begin{figure}[!b]%figure1
\centerline{\includegraphics[scale=0.42]{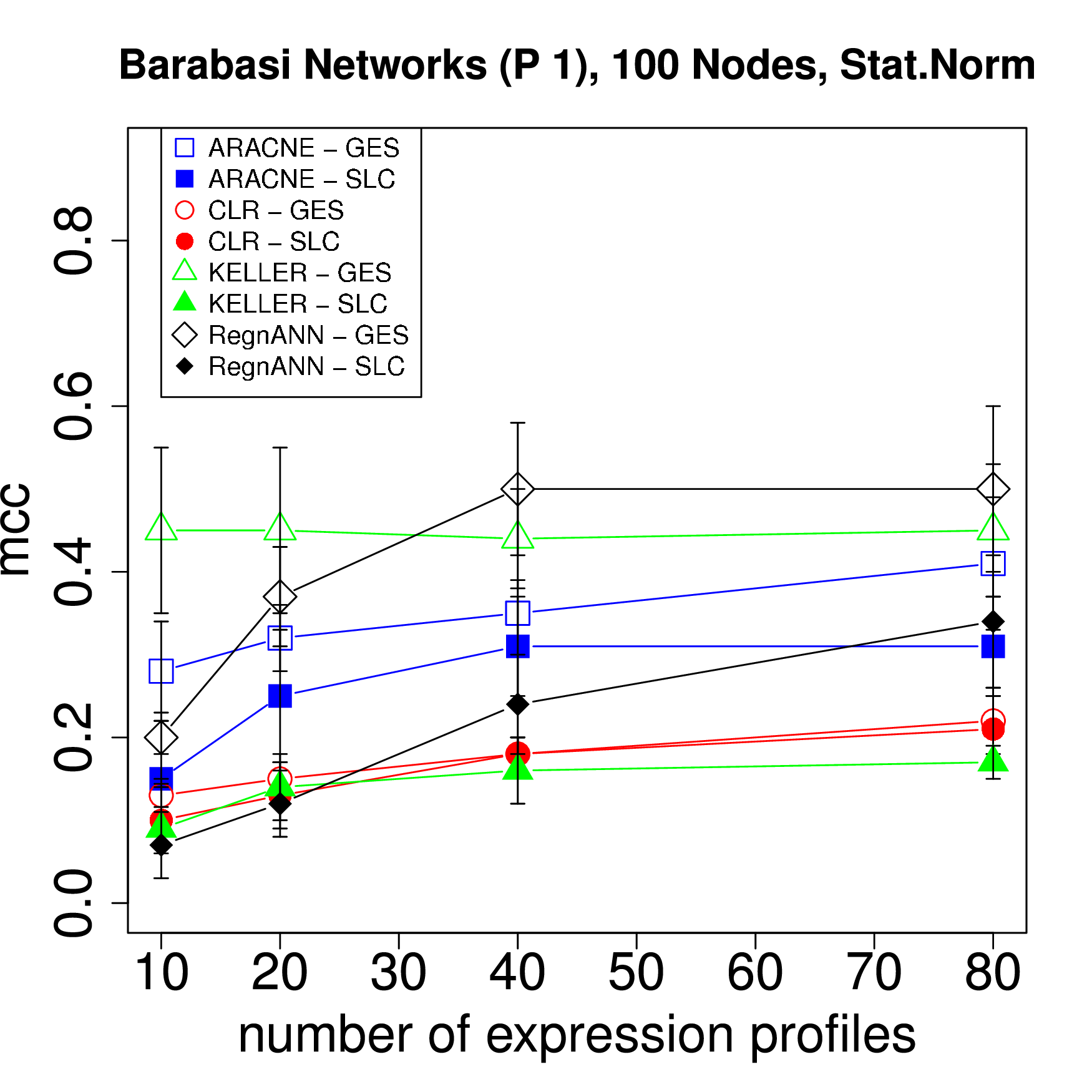}}
\caption{MCC scores of the different network inference algorithms for synthetic Barabasi networks (scale free, exponent $P=1$), varying number of expression profiles and constant number of nodes ($100$). Both methods (GES and SLC) for data synthesis are considered. Expression values are statistically normalized (zero mean and unit standard deviation).}
\label{fig:Barabasi_mcc_vs_data}
\end{figure}

Figure \ref{fig:Barabasi_mcc_vs_data} shows the MCC scores for the same network inference methods as above, varying the number of expression profiles considered while keeping constant the size of the Barabasi network ($100$ nodes). Expression values are statistically normalized.
%
%%%%%%%%%%%%%%%%%%%%%%%%%%%%%%%%%%%%%%%%%%%
%%%%%%%%%%%%%%%%%%%%%%%%%%%%%%%%%%%%%%%%%%%
%
Figure \ref{fig:Barabasi_mcc_vs_data} indicates that the MCC scores
greatly vary when considering statistically normalized values while
varying the amount of data generated (the number of expression
profiles). The data synthesis method adopted can also greatly affect
the performance score. MCC scores for RegnANN, ARACNE and CLR show to
be positively affected when the number of generated expression
profiles is increased from 10 to 40: RegnANN-GES scores $0.20 \pm 0.02$
considering only 10 profiles, while scoring $0.50 \pm 0.08$ with 40 different. 
Adopting SLC data synthesis, RegnANN scores $0.07 \pm 0.04$ and $0.24 \pm 0.06$
with 10 and 40 expression profiles respectively. Similarly, ARACNE-GES
scores $0.28 \pm 0.06$ and $0.35 \pm 0.04$ with 10 and 40 expression
profiles respectively. ARACNE-SLC scores $0.15 \pm 0.08$ and $0.31 \pm 0.06$ with 10
and 40 expression profiles respectively. On the other hand, as also
shown in Figure \ref{fig:Barabasi_mcc_vs_nodes}, CLR shows performance
curves that are not influenced by the data synthesis method adopted:
it scores $0.13 \pm 0.01$ (GES) with 10 expression profiles; $0.10 \pm
0.04$ synthesizing data with SLC. With 40 expression profiles CLR-GES
scores $0.22 \pm 0.04$;CLR-SLC scores $0.21 \pm 0.04$. On
the contrary, Figure \ref{fig:Barabasi_mcc_vs_data} shows that the
performance of KELLER is greatly influenced by the data synthesis
method, while the number of expression profiles has a somewhat limited
impact: KELLER scores $0.44 \pm 0.06$ synthesizing expression profiles
with GES ($40$ in total), it scores $0.18 \pm 0.02$ using SLC to
generate $40$ profiles.

\begin{figure}[!t]
\centerline{\includegraphics[scale=0.40]{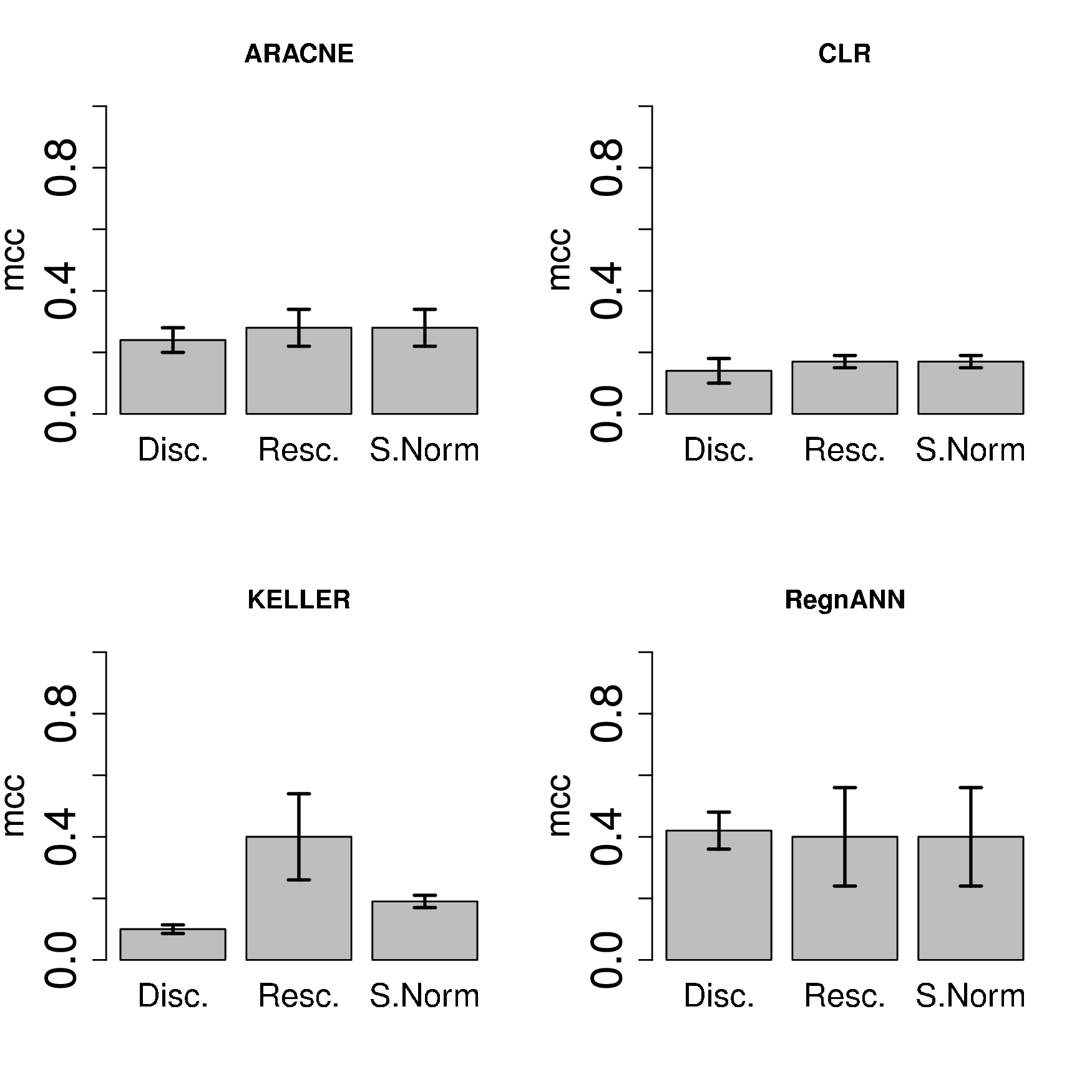}}
\caption{MCC scores of the different network inference algorithms for synthetic Barabasi networks (scale free, exponent $P=1$), varying data normalization method (Discretization, Linear Rescaling and Statistical Normalization) and constant network size ($200$ nodes) and number of expression profiles generated ($160$). Only SLC data synthesis is considered.}
\label{fig:Barabasi_mcc_vs_norm}
\end{figure}

Figure \ref{fig:Barabasi_mcc_vs_norm} shows the MCC scores obtained with ARACNE, CLR, KELLER and RegnANN by varying data normalization methods while keeping constant the network size ($200$ nodes) and the number of expression profiles generated ($160$). Only the SLC data synthesis is considered.
%
%%%%%%%%%%%%%%%%%%%%%%%%%%%%%%%%%%%%%%%%%%%
%%%%%%%%%%%%%%%%%%%%%%%%%%%%%%%%%%%%%%%%%%%
%
Figure \ref{fig:Barabasi_mcc_vs_norm} indicates that ARACNE, CLR and
RegnANN MCC scores are not significantly affected -- considering the
error of the measure -- by the normalization method: RegnANN scores
$0.42 \pm 0.06$, $0.4 \pm 0.1$ and $0.4 \pm 0.1$ applying respectively
discretization, linear rescaling and statistical normalization to the
data. Similarly, ARACNE scores $0.24 \pm 0.04$, $0.28 \pm 0.03$ and
$0.28 \pm 0.03$ when the expression values are discretized, linearly
rescaled and statistically normalized. Finally, CLR scores: $0.14 \pm
0.04$, $0.17 \pm 0.01$ and $0.17 \pm 0.01$ for the very same
normalization methods above (discretization, linear rescaling,
statistical normalization). On the other hand, KELLER MCC scores show
to be highly influenced by the normalization method applied to the
synthetic data. In the case of discretization and in the case of statistical normalization KELLER scores $0.10 \pm 0.01$ and $0.19
\pm 0.01$ respectively. In the case of linear rescaling it scores a higher value: $0.40 \pm 0.07$.

\begin{figure}[!t]%figure1
\centerline{\includegraphics[scale=0.42]{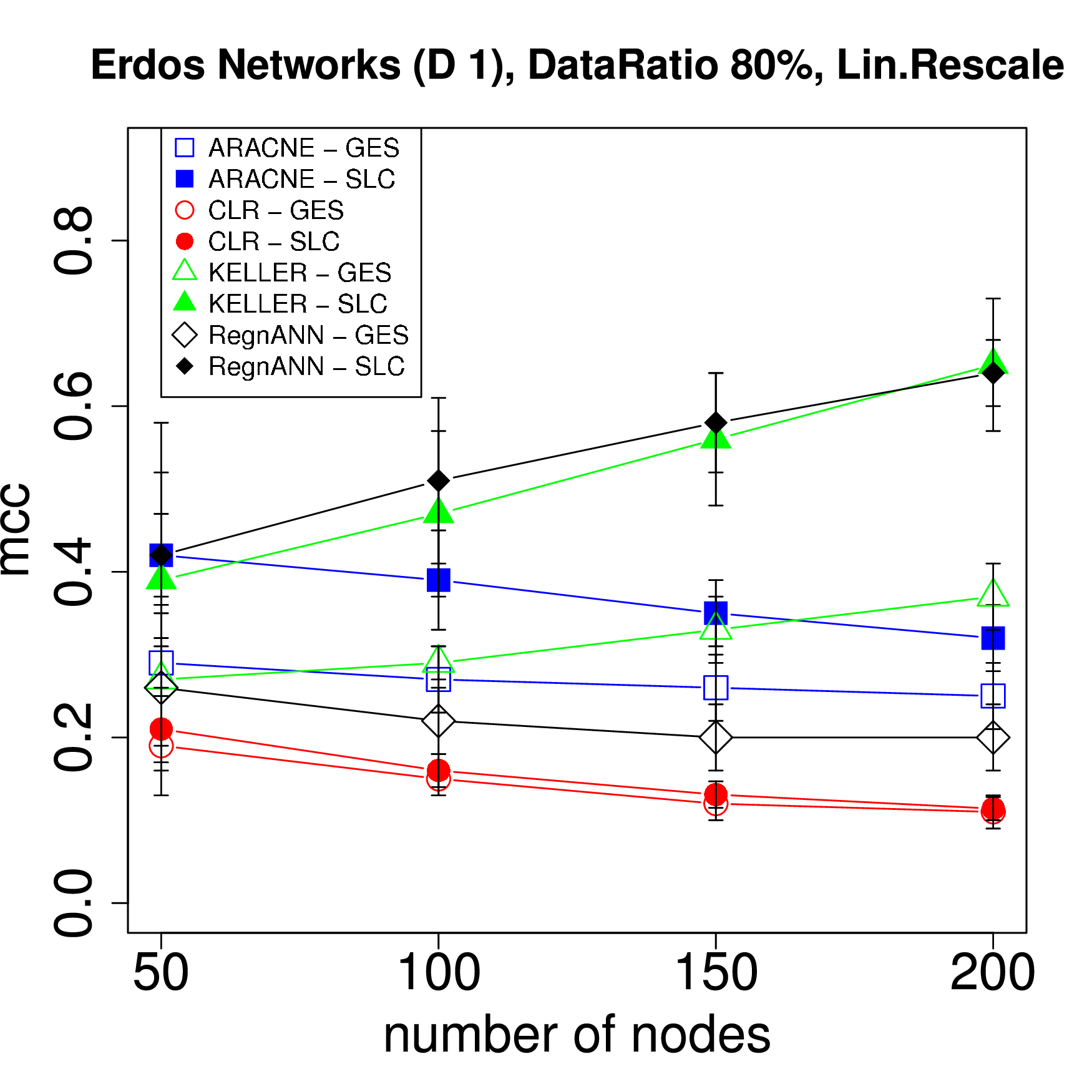}}
\caption{MCC scores of the different network inference algorithms for synthetic Erd\"os-R\'enyi networks (random graph, mean degree $D=1$), varying the number of nodes and keeping constant the data ratio: the number of expression profiles to the number of nodes ($80\%$). Both methods (GES and SLC) for data synthesis are considered. Expression values are linearly rescaled in $[-1,1]$.}
\label{fig:Erdos_mcc_vs_nodes}
\end{figure}

Figure \ref{fig:Erdos_mcc_vs_nodes} shows the MCC scores obtained with ARACNE, CLR, KELLER and RegnANN for synthetic Erd\"os-R\'enyi networks (random graph, mean degree $D=1$), varying the number of nodes. In order to provide similar amount of information to the inference algorithms while varying the size of the network, we kept constant the data ratio ($80\%$). Expression values are linearly rescaled in $[-1,1]$.
%
%%%%%%%%%%%%%%%%%%%%%%%%%%%%%%%%%%%%%%%%%%%
%%%%%%%%%%%%%%%%%%%%%%%%%%%%%%%%%%%%%%%%%%%
%
In the case of Erd\"os-R\'enyi networks the MCC curves are greatly and unevenly
affected by all the parameters explored: inference method, size of the
network and data synthesis method. ARACNE and CLR show a decreasing MCC
score -- although not strictly statistically significant -- when the
number of nodes in the network is increased from $50$ to $200$: ARACNE-GES
scores $0.29 \pm 0.08$ with network size $50$, $0.25 \pm 0.04$
with network size $200$. Similarly, CLR-GES scores $0.19 \pm 0.06$
with network size $50$, $0.11 \pm 0.02$ with network size
$200$ -- a similar negative trend is recorded in case of SLC data
synthesis. On the other hand, KELLER and RegnANN have higher MCC
when the number of nodes in the network is increased from
$50$ to $200$: KELLER-SLC scores $0.39 \pm 0.08$ network size
$50$, $0.65 \pm 0.08$ when the network size is $200$. Similarly,
RegnANN-SLC scores $0.4 \pm 0.1$ and $0.64 \pm 0.04$ for
network size $50$ and $200$ respectively. Considering GES for
synthetic data generation, the MCC curves are significantly different
for both KELLER and RegnANN: KELLER scores $0.37 \pm 0.04$ for network
size $200$ while RegnANN scores $0.20 \pm 0.04$ for similarly sized
networks ($200$ nodes).

\begin{figure}[!t]%figure1
\centerline{\includegraphics[scale=0.42]{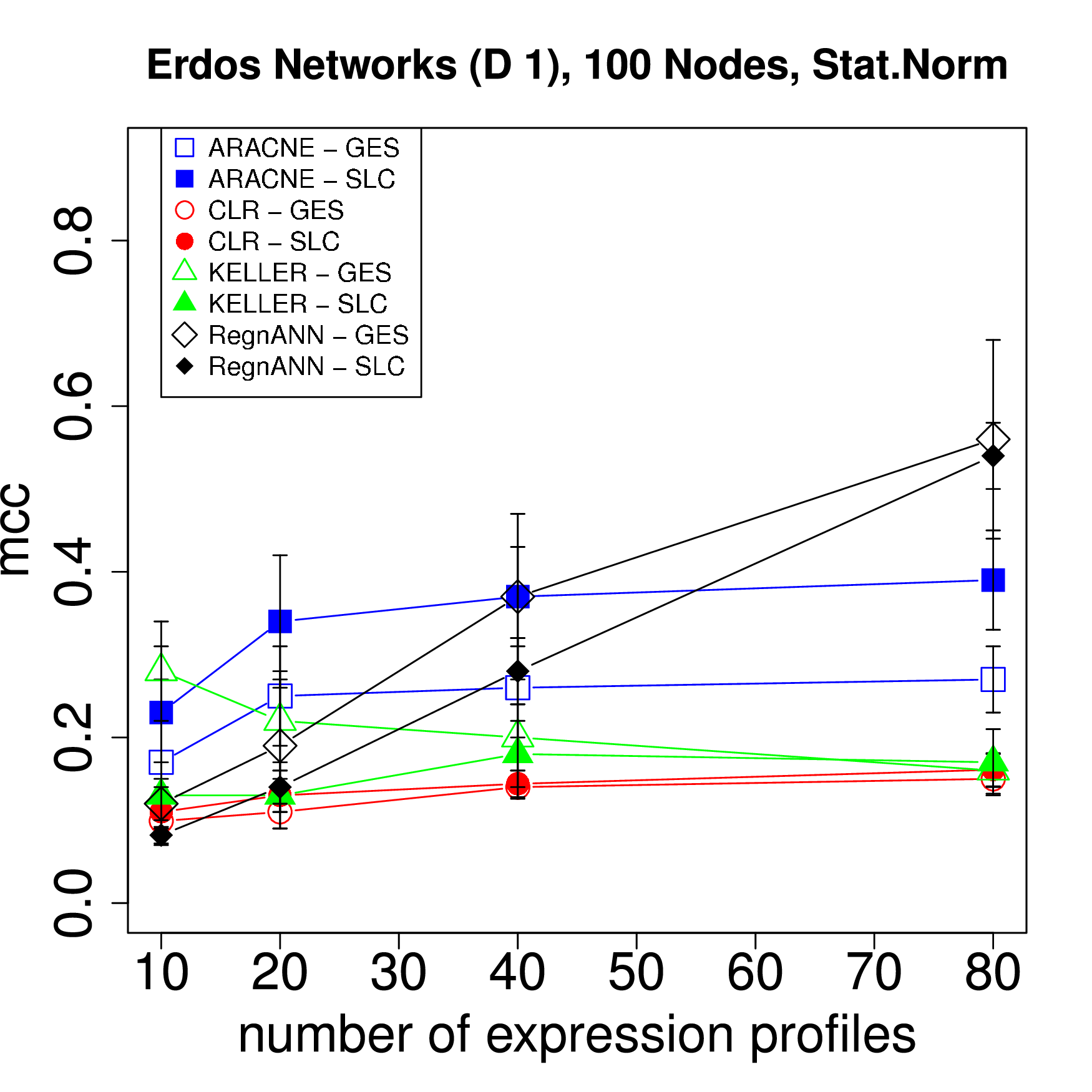}}
\caption{MCC scores of the different network inference algorithms for synthetic Erd\"os-R\'enyi networks (random graph, mean degree $D=1$), varying the number of expression profiles and keeping constant the number of nodes ($100$). Both methods (GES and SLC) for data synthesis are considered. Expression values are statistically normalized (zero mean and unit standard deviation).}
\label{fig:Erdos_mcc_vs_data}
\end{figure}

Figure \ref{fig:Erdos_mcc_vs_data} shows the MCC scores for the same network inference methods as above, varying the number of expression profiles considered while keeping constant the size of the Erd\"os-R\'enyi network ($100$ nodes). Expression values are statistically normalized.
%
%%%%%%%%%%%%%%%%%%%%%%%%%%%%%%%%%%%%%%%%%%%
%%%%%%%%%%%%%%%%%%%%%%%%%%%%%%%%%%%%%%%%%%%
%
As indicated in Figure \ref{fig:Erdos_mcc_vs_data}, KELLER and RegnANN show opposite MCC curves by increasing the amount of expression profiles generated. RegnANN-GES shows rapidly increasing scores varying the number of expression profiles from $10$ to $80$: $0.12 \pm 0.02$ and $0.6 \pm 0.1$ respectively. KELLER-GES scores $0.28 \pm 0.06$ and KELLER-SLC scores $0.13 \pm 0.04$ with 10 expression profiles. KELLER-GES scores $0.16 \pm 0.01$ and KELLER-SLC scores $0.17 \pm 0.04$ with $80$ expression profiles. On the other hand, MCC curves for CLR are limitedly affected by the number of expression profiles or by the data generation methodology: with $80$ expression profiles it scores $0.15 \pm 0.02$ using GES and $0.16 \pm 0.02$ using SLC for data synthesis.

\begin{figure}[!t]
\centerline{\includegraphics[scale=0.40]{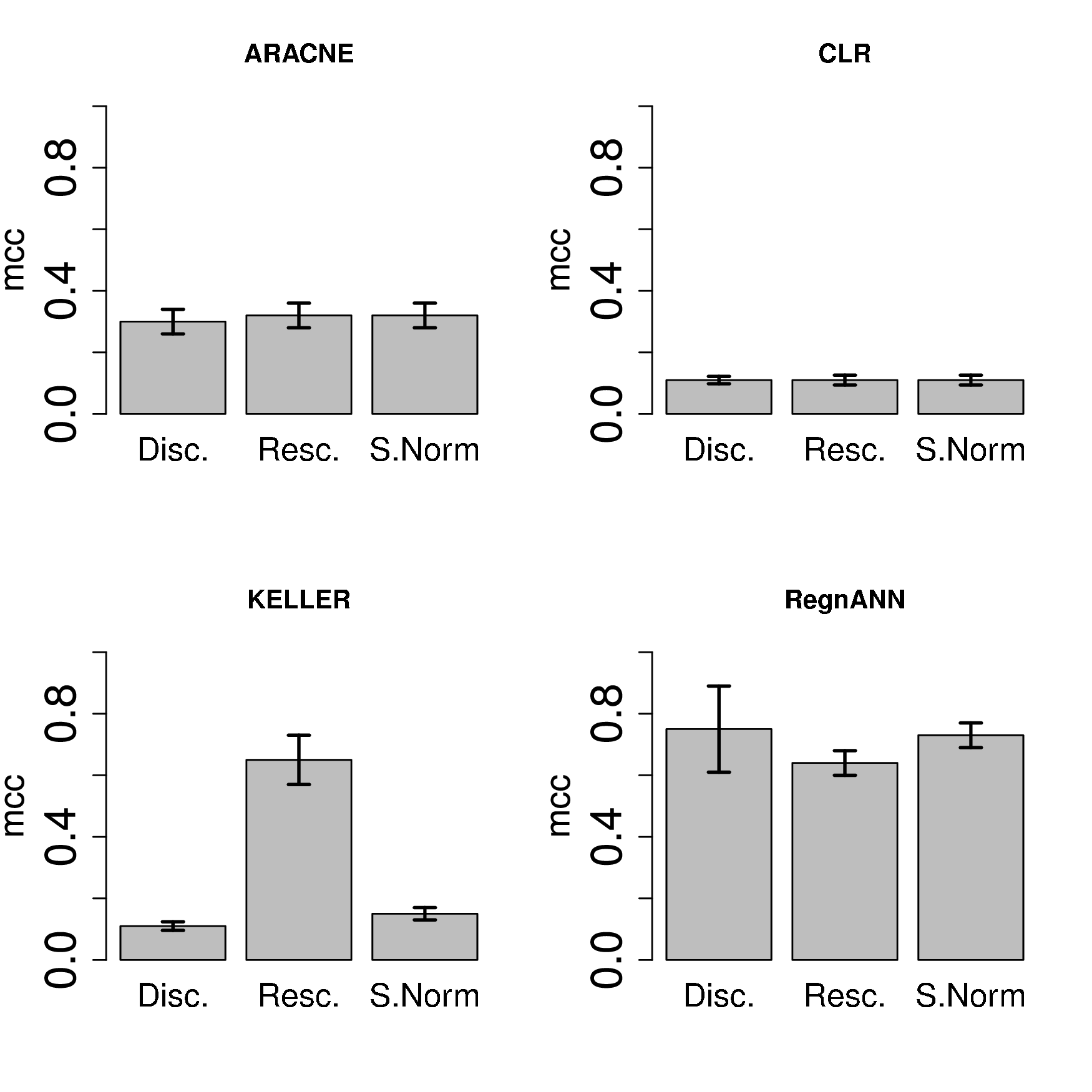}}
\caption{MCC scores of the different network inference algorithms for synthetic Erd\"os-R\'enyi networks (random graph, mean degree $D=1$), varying data normalization method (Discretization, Linear Rescaling and Statistical Normalization) and constant network size ($200$ nodes) and number of expression profiles generated ($160$). Only SLC data synthesis is considered.}
\label{fig:Erdos_mcc_vs_norm}
\end{figure}

Figure \ref{fig:Erdos_mcc_vs_norm} shows the MCC scores obtained with ARACNE, CLR, KELLER and RegnANN varying data normalization method while keeping constant the network size ($200$ nodes) and the number of expression profiles generated ($160$). Only the SLC data synthesis is considered.
%
%%%%%%%%%%%%%%%%%%%%%%%%%%%%%%%%%%%%%%%%%%%
%%%%%%%%%%%%%%%%%%%%%%%%%%%%%%%%%%%%%%%%%%%
%
As in the case of Barabasi networks (Figure \ref{fig:Barabasi_mcc_vs_norm}), Figure \ref{fig:Erdos_mcc_vs_norm} shows that ARACNE, CLR and RegnANN MCC scores are not significantly affected by the normalization method. On the contrary, KELLER is significantly affected: it scores $0.11 \pm 0.01$ and  $0.15 \pm 0.02$ when the expression values are discretized and statistically normalized respectively. A higher value of  $0.65 \pm 0.08$ is recorded in the linearly rescaled case.

\subsubsection*{Selected Escherichia coli subnetworks:}
\label{ssec:ecoli}
Table \ref{Tab:e.coli} summarizes the results obtained on a selection of \emph{Escherichia coli} gene subnetworks (\cite{alvarez09modular}) for the four inference algorithms. Gene expression values are linearly rescaled in $[-1,1]$.

%%%%%%%%%%%%%%%%%%%%%%%%%%%%%%%%%%%%%%%%%%
\begin{table}[!t]
\processtable{MCC scores of the different network inference algorithms on selected \emph{Escherichia coli} network modules\label{Tab:e.coli}}
{\begin{tabular}{c c r r r r r r r}\toprule
ID & D & NN & NL & A. & CLR & K. & R.ANN & Err\\\midrule
81 & 0.245 & 7 		& 12	& 0.78 & 0.45 & -0.12 & 0.4 & 0.1\\
6  & 0.189 & 13 	& 32	& 0.13 & 0.29 & 0.02 & 0.3 & 0.1\\
12 & 0.180 & 10 	& 18	& 0.43 & 0.42 & 0.63& 0.32 & 0.00\\
75 & 0.133 & 16 	& 34	& 0.10 & 0.24 & 0.10 & 0.23 & 0.08\\
88 & 0.100 & 19 	& 36	& 0.00 & 0.17 & -0.07 & -0.05 & 0.02\\
96 & 0.001 & 104 	& 18	& 0.08 & 0.02 & 0.00& 0.00 & 0.01\\
94 & 0.000 & 81 	& 2		& 0.09 & 0.02 & 0.15 & 0.026  & 0.001\\\botrule

\end{tabular}}{Column ID indicates the id of the network module as in \cite{alvarez09modular}, D the density of the module (the ratio of the number of links to the square of the number of nodes), NN the number of nodes in the module, LN the number of links. Column A. shows results for ARACNE, Column K. results for KELLER and R.ANN the results for RegnANN.  Column Err refers to the error associated to the MCC score of RegnANN and it is calculated as twice the standard deviation of $10$ independent runs.}
\end{table}
%%%%%%%%%%%%%%%%%%%%%%%%%%%%%%%%%%%%%%%%%%
%
As for the case of synthetic data, Table \ref{Tab:e.coli} indicates
great variability of the MCC scores across the different network
modules for all the inference methods tested. ARACNE scores range from
$0.78$ (module $81$) to $0.00$ (module $88$). CLR values range between
$0.45$ and $0.02$ for module $81$ and $96$ respectively. KELLER scores
range between $0.63$ and $-0.12$ (module $12$ and module $81$
respectively). Finally RegnANN scores range between $0.32 \pm 0.00$
\footnote{In this case the error associated to the measure is 0: the
very same result is obtained for all repetitions.} (module $12$) and
$-0.05 \pm 0.02$ (module $88$). It is interesting
to note that the MCC score varies unevenly for the different inference
algorithms with respect to the module network density (the ratio of
the number of links to the square of the number of nodes), e.g.:
ARACNE scores $0.13$ on module $6$ (density $D=0.189$) and scores
$0.43$ on module $12$ (density $D=0.189$). On the same two modules,
CLR scores $0.29$ and $0.39$ respectively while KELLER scores $0.02$
and $0.63$. On the other hand, RegnANN are more homogeneous: it scores
$0.3 \pm 0.1$ and $0.32 \pm 0.00$ on module $6$ and module $12$
respectively.

\section{Discussion}
\label{sec:discussion}
The analysis of the results obtained on synthetic data shows that the
performance of the inference methods are highly and unevenly
influenced by the simulation parameters: the topology of the network
and its size, the method used to synthesize the expression values and
the raw data normalization step adopted. The dependency of the results
on all the parameters of the simulations makes objectively very hard
the task of establishing which method performs best. Generally,
RegnANN shows performance scores that compare very favorably with all
the other inference method tested. The solution based on ANN provides
good results on both Barabasi and Erd\"os-R\'enyi networks varying the
number of expression profiles synthesized (Figure
\ref{fig:Barabasi_mcc_vs_data} and Figure
\ref{fig:Erdos_mcc_vs_data}), and it shows stable MCC scores with
regards to the different data normalization adopted (Figure
\ref{fig:Barabasi_mcc_vs_norm} and Figure
\ref{fig:Erdos_mcc_vs_norm}).  The evaluation on synthetic data
indicates that CLR is the most stable inference method with regards to
variations in the network topology and in the data synthesis, although
it shows MCC scores that compare unfavorably with the other
methods. On the other hand, ARACNE compares favorably with KELLER and
RegnANN in terms of MCC, showing also stability with regards to the
different data normalization adopted (Figure
\ref{fig:Barabasi_mcc_vs_norm} and Figure
\ref{fig:Erdos_mcc_vs_norm}). On the contrary, KELLER shows a great
deal of variability in the MCC scores with regards to the different
data normalization methods: Figure \ref{fig:Barabasi_mcc_vs_norm} and
Figure \ref{fig:Erdos_mcc_vs_norm} indicate that this algorithm
performs best when the expression profiles are linearly rescaled in
$[-1,1]$. These results suggest that in \cite{song09keller}, the
algorithm may be not be performing at its best since the author
discretized the expression values in $\{-1, 1\}$.

The results on the \emph{Escherichia coli} gene network modules confirm that
the inference algorithms tested show great variability in the MCC
scores, suggesting that the correctness of the inferred network depends on the
topological properties of the modules (the very same expression values
are used to infer the different gene sub-networks), in accordance to findings in \cite{altay10revealing}.

The great deal of variability in the results for both synthetic and
real-world data indicates that each inference method can potentially
select the correct network topology -- or the incorrect one --
depending on a number of factors which may not be limited to the
relative small set of parameters explored here. 
With regard to this, we lastly verify possible stability issues of the network
inference algorithms related to the re-generation of synthetic data
for a given sample network topology. 
Table \ref{Tab:stability} shows
the results obtained on a sample Barabasi network and a sample
Erd\"os-R\'enyi network (fixed topology, 100 nodes data ratio $80\%$)
by applying each inference algorithm $10$ times on $10$ different
simulated expression values (SLC method). Column Accuracy indicates
the mean MCC score for the $10$ inferred adjacency matrices with
respect to the ground-truth (A.Err is the associated error calculated
as twice the standard deviation of the mean). Column Stability
indicates the mean distance among all the inferred topologies: a value
equal to 1 indicates perfect stability, e.g.: the same topology is
reconstructed all the times (similarly, $0$ indicates random results)
-- S.Err is the associated error calculated as twice the standard
deviation of the mean.
%
%%%%%%%%%%%%%%%%%%%%%%%%%%%%%%%%%%%%%%%%%%
\begin{table}[!t]
\processtable{Accuracy [MCC] scores in network topology inference for the different reverse engineering algorithms and their stability [MCC]\label{Tab:stability}.}
{\begin{tabular}{l c r c r}
\toprule
\multicolumn{3}{l}{Synthetic Barabasi Network} &  &\\
        & Accuracy [MCC] & A.Err & Stability [MCC] & S.Err\\\midrule
ARACNE  & 0.31 & 0.04 & 0.21 & 0.04\\
CLR     & 0.23 & 0.02 & 0.24 & 0.03\\
KELLER  & 0.40 & 0.07 & 0.58 & 0.06\\
RegnANN & 0.37 & 0.06 & 0.42 & 0.07\\\midrule

\multicolumn{3}{l}{Synthetic Erd\"os-R\'enyi Network} &  &\\
        & Accuracy [MCC] & A.Err & Stability [MCC] & S.Err\\\midrule
ARACNE  & 0.39 & 0.02 & 0.17 & 0.03\\
CLR     & 0.16 & 0.01 & 0.07 & 0.03\\
KELLER  & 0.47 & 0.07 & 0.27 & 0.05\\
RegnANN & 0.49 & 0.06 & 0.28 & 0.05\\\botrule
\end{tabular}}{Column Accuracy indicates the mean MCC score in reconstructing the target network topology, column A.Err the associated error. Column Stability indicates the mean distance [MCC] among all the inferred topologies, column S.Err the associated error.}
\end{table}
%%%%%%%%%%%%%%%%%%%%%%%%%%%%%%%%%%%%%%%%%%
%
Table \ref{Tab:stability} suggests again that all the methods suffer of problems related to the variability of the inferred network topology: no method shows a stability score close to $1$. KELLER scores best on the Barabasi network (stability of $0.58 \pm 0.06$, accuracy $0.40 \pm 0.07$). Both KELLER and RegnANN score best on the Erd\"os-R\'enyi network (a stability of about $0.27$, an accuracy of about $0.47$).

\section{Conclusion}
\label{sec:conclusion}
In this work we presented a novel method for network inference based on an ensemble of multi-layer perceptrons configured as multi-variable regressor (RegnANN). We compared its performance to the performance of three different network inference algorithms (ARACNE, CLR and KELLER) on the task of reverse engineering the gene network topology, in terms of the associated MCC score. 

Our extensive evaluation indicates that all the algorithms suffer of instability in the reconstruction of the network topology due to the various sources of variability, possibliy not limited to the relative small set of parameters explored here. Because of such instability, it is objectively very difficult to establish which method performs best. Generally, the newly introduced RegnANN shows performance scores that compare very favorably with all the other inference methods tested. Nonetheless further efforts are required in order to effectively cope with the difficulty of the task and minimize the variability of the inference process.
%
%The great deal of variability in the results for both synthetic and
%real-world data indicates that each inference method can potentially
%select the correct network topology -- or the incorrect one --
%depending on a number of factors which may not be limited to the
%relative small set of parameters explored in here. 

\section*{Acknowledgements}
The author want to thank Cesare Furlanello for his precious remarks and for reading earlier versions of the paper.
\paragraph{Funding} 
This work is funded by the European Union FP7 Project HiperDART and by the Italian Ministry of Health Project ISITAD (RF2007 conv.42).
\bibliographystyle{natbib}

\bibliography{regnann_grimaldi}
\include{rgn_suppmat}
\end{document}

%% file: rgn_suppmat.tex
\appendix
\section*{Appendix}
\section{Gene Expression Normalization}
\label{ssec:normalization}
Generally, in microarray experiments, the  analysis of the raw data is often hampered by a number of technical and statistical problems. The possible remedies usually lie in appropriate preprocessing steps, proper normalization of the data and application of statistical testing procedures in the derivation of differentially expressed genes (\cite{steinhoff06normalization}). Although many of the real-world issues in data preprocessing and normalization do not apply here, we are interested in verifying how discretization and rescaling -- some of the most common (and possibly simple) steps taken to normalize the raw data -- can impact the accuracy of the network inference algorithms here considered.
 
\subsection{Discretization}  
It is often the case that a number of sources of noise can be introduced into the microarray measurements, e.g. during the stage of hybridization, digitization and normalization. Therefore, it is often preferred to consider only the qualitative level of gene expression rather than its actual value (\cite{song09keller}):  gene expression is modeled as either being up-regulated ($+1$) or down-regulated ($-1$) by comparing the given value to a threshold. For example, in \cite{tuna09cross} it is shown that binarizing gene expression data leads to classification outcomes very similar to the results obtained on real-valued data.

In this work we compute the discrete value of the expression for each of the $N$ genes at each of the $M$ steps as the sign of the difference of the expression values of the given gene at step $m$ and step $m-1$.

\subsection{Rescaling} 
Generally, when a scaling method is applied to the data, it is assumed that different sets of intensities differ by a constant global factor (\cite{steinhoff06normalization}). It may also happen that the rescaling is a necessary step due to the inference method adopted, as in the case of SVM (Support Vector Machine) or ANN (Artificial Neural Network) classification/regression. 

In this work we test two different data rescaling methods:

\begin {itemize}
	\item \emph{linear rescaling}: each gene expression column-vector is linearly  rescaled between $\left[-1,1\right]$;
	\item \emph{statistical normalization}: each gene expression column-vector\footnote{In this work we consider gene expression matrices of dimension $M \times N$: $N$ genes whose expression levels are recorded $M$ times.} is rescaled such that its mean value is equal to $0$ and the standard deviation equal to $1$.
\end {itemize}

\section{Performance metric}
\label{ssec:performance}
When the performance of a network inference method is evaluated, it is common practice to adopt two metrics: precision and recall. Recall indicates the fraction of true interactions correctly inferred by the algorithm, and is estimated according to the following equation:

\begin{equation}
Recall = \frac{TP}{TP + FN}
\end{equation}

where TP indicates the fraction of \emph{true positives}, while FN indicates the fraction of \emph{false negatives}.

On the other hand, precision measures the fraction of true interactions among all inferred ones, and it is computed as:

\begin{equation}
Precision = \frac{TP}{TP + FP}
\end{equation}

where FP indicates the ratio of \emph{false positives}.

In this work we adopt instead the Matthews correlation coefficients -- MCC (\cite{baldi00assessing, matthews75comparison}): this is a measure that takes into account both true/false positives and true/false negatives and it is generally regarded to as a balanced measure, useful specially in the case of unbalanced classes (i.e.: not equal number of positive and negative examples). 

The MCC is in essence a correlation coefficient between the observed and predicted binary classifications: it returns a value between $-1$ and $+1$. A coefficient value equal to $+1$ represents a perfect prediction, $0$ indicates an average random prediction while $-1$ an inverse prediction (\cite{baldi00assessing, matthews75comparison}). In the context of network topology inference the observed class is the true network adjacency matrix, while the predicted class is the inferred one. 

The Matthews correlation coefficient has the following is obtained according to the following equation:
\begin{equation}
\textrm{MCC} = \frac{\textrm{TP}\cdot\textrm{TN}-\textrm{FP}\cdot\textrm{FN}}{\sqrt{\left(\textrm{TP}+\textrm{FP}\right)\left(\textrm{TP}+\textrm{FN}\right)\left(\textrm{TN}+\textrm{FP}\right)\left(\textrm{TN}+\textrm{FN}\right)}}\ .
\end{equation}
Recently MCC has also been used for comparing network topologies (\cite{supper07reconstructing, stokic09fast}).

\section{Synthetic Data Generation}
The synthetic data sets used in the main paper are obtained starting from an adjacency matrix describing the selected topology\footnote{The network graph is generated using the \emph{igraph} extension package to the GNU R project for Statistical Computing.}. In this work we consider undirected graphs: we are interested in estimating the structures of interaction between nodes/genes, rather than the detailed strength or the direction of these interactions. Thus, we consider only symmetric and discrete adjacency matrices, representing with a value of $1$ the presence of a link between two nodes. A value equal to $0$ in the adjacency matrix indicates no interaction. 

\subsection{Network Topology}
\begin{figure*}[!b]
\centerline{\includegraphics[scale=0.55]{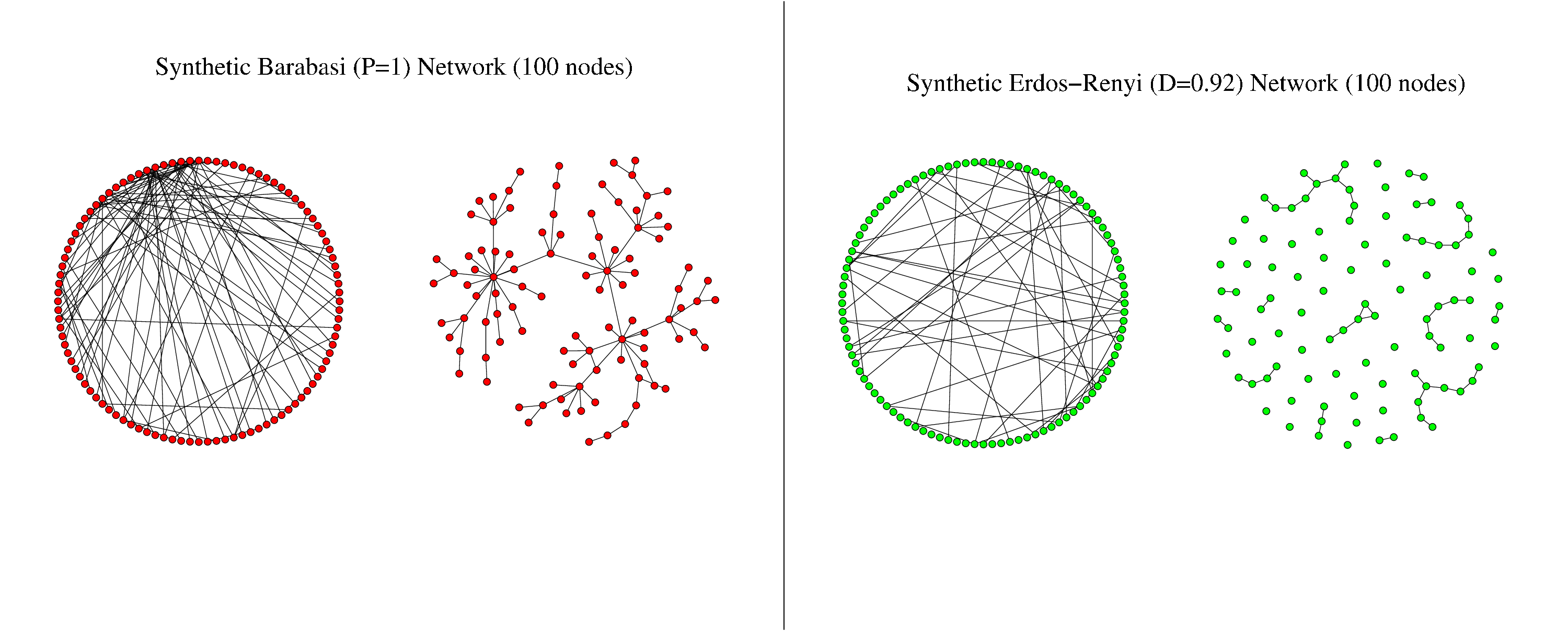}} 
\caption{Sample network topologies: left, Barabasi Network with 100 nodes (power-law exponent $P$ equal to 1); right Erd\"os-R\'enyi network, 100 nodes and average degree ($D$) egual to $0.92$.  }
\label{fig:topology}
\end{figure*}

Here we consider two different network topologies: Barabasi-Albert (\cite{barabasi99emergence}) and Erd\"os-R\'enyi (\cite{erdos59random}).
Figure \ref{fig:topology} shows two sample network topologies: left, Barabasi Network with 100 nodes (power-law exponent $P$ equal to 1); right, Erd\"os-R\'enyi network, 100 nodes and average degree ($D$) equal to $0.92$.

%%%%%%%%%%%%%%%%%%%%%%%%%%%%%%%%%%%%%%%%%%%
%%%%%%%%%%%%%%%%%%%%%%%%%%%%%%%%%%%%%%%%%%%

Once the topology of the network is (randomly) generated, the output profiles of each node are generated according to the approaches in the following section.

\subsection{Gene Expression Synthesis}

\subsubsection{Simple Linear Correlation (SLC):}
similarly to the simulation of gene expression data presented in the supplementary material of \cite{langfelder07eigengene}, we consider a set of \emph{seed} expressions (a matrix $M \times N$ -- $N$ genes which expression profiles are recoded $M$ times -- with values uniformly distributed in [-1, 1]) and the desired topology expressed by the adjacency matrix $adjM$ ($N \times N$). The gene expression profiles ($gep$, a matrix $M \times N$) are calculated as:

\begin{equation}
	gep = seed + seed \star adjM
\end{equation}

where the symbol `$+$' indicates element-element summation and the symbol `$\star$' indicates row-column matrix multiplication. With this method, the \emph{seed} expression columns are linearly correlated (correlation equal to $1$) with the columns of the same matrix as described by the discrete input adjacency matrix \emph{adjM}.

\subsubsection{Gene Expression Simulator (GES):} 
this second methodology is based on a gene network simulator recently proposed to assess reverse engineering algorithms (\cite{dicamillo09gene}). Given an input adjacency matrix, the network simulator uses fuzzy logic to represent interactions among the regulators of each gene and adopts differential equations to generate continuous data. As in \cite{margolin06aracne}, we obtain synthetic expression values of each gene $n$ ($n = 1,\ldots,N$) by simulating its dynamics until the expression value reaches its steady state. We obtain $M$ different values for each gene by repeating the process $M$ times and recording the expression value at steady state. The synthesis of each gene profile is randomly initialized by the simulator.

%% file: regnann_grimaldi.bbl
\begin{thebibliography}{}

\bibitem[Altay and Emmert-Streib(2010)Altay and
  Emmert-Streib]{altay10revealing}
Altay, G. and Emmert-Streib, F. (2010).
\newblock {Revealing differences in gene network inference algorithms on the
  network level by ensemble methods}.
\newblock {\em Bioinformatics\/}, {\bf 26}(14), 1738--1744.

\bibitem[Baldi {\em et~al.}(2000)Baldi, Brunak, Chauvin, Andersen, and
  Nielsen]{baldi00assessing}
Baldi, P., Brunak, S., Chauvin, Y., Andersen, C., and Nielsen, H. (2000).
\newblock {Assessing the accuracy of prediction algorithms for classification:
  an overview}.
\newblock {\em Bioinformatics\/}, {\bf 16}(5), 412--424.

\bibitem[Bansal {\em et~al.}(2007)Bansal, Ambesi-Impiombato, and
  di~Bernardo]{bansal07how}
Bansal, M.~Belcastro, V., Ambesi-Impiombato, A., and di~Bernardo, D. (2007).
\newblock {How to infer gene networks from expression profiles}.
\newblock {\em Mol. Syst. Biol.}, {\bf 122}(3), 78.

\bibitem[Barabasi and Albert(1999)Barabasi and Albert]{barabasi99emergence}
Barabasi, A. and Albert, R. (1999).
\newblock Emergence of scaling in random networks.
\newblock {\em Science\/}, {\bf 286}(5439), 509--512.

\bibitem[Baralla {\em et~al.}(2009)Baralla, Mentzen, and de~la
  Fuente]{baralla09inferring}
Baralla, A., Mentzen, W., and de~la Fuente, A. (2009).
\newblock {Inferring Gene Networks: Dream or Nightmare?}
\newblock {\em Ann. N.Y. Acad. Sci.}, {\bf 1158}, 246--256.

\bibitem[Bishop(1995)Bishop]{bishop95neural}
Bishop, C. (1995).
\newblock {\em {Neural Networks for Pattern Recognition}\/}.
\newblock OUP, New York.

\bibitem[Cover and Thomas(1991)Cover and Thomas]{cover91elements}
Cover, T. and Thomas, J. (1991).
\newblock {\em {Elements of Information Theory}\/}.
\newblock Wiley, 2nd edition.

\bibitem[Di~Camillo {\em et~al.}(2009)Di~Camillo, Toffolo, and
  Cobelli]{dicamillo09gene}
Di~Camillo, B., Toffolo, G., and Cobelli, C. (2009).
\newblock {A Gene Network Simulator to Assess Reverse Engineering Algorithms}.
\newblock {\em Ann. N.Y. Acad. Sci.}, {\bf 1158}.

\bibitem[Dimitrova {\em et~al.}(2007)Dimitrova, Jarrah, Laubenbacher, and
  Stigler]{dimitrova07groebner}
Dimitrova, E., Jarrah, A., Laubenbacher, R., and Stigler, B. (2007).
\newblock {A Gr{\"o}bner fan method for biochemical network modeling}.
\newblock In D.~Wang, editor, {\em Proceedings of ISSAC 2007\/}, pages
  122--126.

\bibitem[Erd\"os and Renyi(1959)Erd\"os and Renyi]{erdos59random}
Erd\"os, P. and Renyi, A. (1959).
\newblock {On Random Graphs}.
\newblock {\em Publ. Math. Debrecen\/}, {\bf 6}, 290--297.

\bibitem[Faith {\em et~al.}(2007)Faith, Hayete, Thaden, Mogno, Wierzbowski,
  Cottarel, Kasif, Collins, and Gardner]{faith07large}
Faith, J., Hayete, B., Thaden, J., Mogno, I., Wierzbowski, J., Cottarel, G.,
  Kasif, S., Collins, J., and Gardner, T. (2007).
\newblock {Large-Scale Mapping and Validation of \emph{Escherichia coli}
  Transcriptional Regulation from a Compendium of Expression Profiles}.
\newblock {\em PLoS Biol.}, {\bf 5}(1), e8.

\bibitem[Friedman {\em et~al.}(2000)Friedman, Linial, Nachman, and
  Pe\'{}er]{friedman00using}
Friedman, N., Linial, M., Nachman, I., and Pe\'{}er, D. (2000).
\newblock {Using Bayesian networks to analyze expression data}.
\newblock {\em J. Comput. Biol.}, {\bf 7}, 601--620.

\bibitem[Glass and Kauffman(1973)Glass and Kauffman]{glass73logical}
Glass, L. and Kauffman, S. (1973).
\newblock {The logical analysis of continuous, non-linear biochemical control
  networks}.
\newblock {\em J. Theor. Biol.}, {\bf 39}, 103--129.

\bibitem[He {\em et~al.}(2009)He, Balling, and Zeng]{he09reverse}
He, F., Balling, R., and Zeng, A.-P. (2009).
\newblock {Reverse engineering and verification of gene networks: Principles,
  assumptions, and limitations of present methods and future perspectives}.
\newblock {\em J. Biotechnol.}, {\bf 144}(3), 190--203.

\bibitem[Karlebach and Shamir(2008)Karlebach and Shamir]{karlebach08modelling}
Karlebach, G. and Shamir, R. (2008).
\newblock Modelling and analysis of gene regulatory networks.
\newblock {\em Nat. Rev. Mol. Cell Biol.}, {\bf 9}, 770--780.

\bibitem[Kauffman(1993)Kauffman]{kauffman93origins}
Kauffman, S. (1993).
\newblock {\em {The Origins of Order: Self-Organization And Selection in
  Evolution}\/}.
\newblock OUP, Oxford.

\bibitem[Krishnan {\em et~al.}(2007)Krishnan, Giuliani, and
  Tomita]{krishnan07indeterminacy}
Krishnan, A., Giuliani, A., and Tomita, N. (2007).
\newblock Indeterminacy of reverse engineering of gene regulatory networks: The
  curse of gene elasticity.
\newblock {\em PLoS ONE\/}, {\bf 2}(6), e562.

\bibitem[Lahabar {\em et~al.}(2008)Lahabar, Agrawal, and
  Narayanan]{lahabar08high}
Lahabar, S., Agrawal, P., and Narayanan, P. (2008).
\newblock {High Performance Pattern Recognition on GPU}.
\newblock In {\em Proceedings of NCVPRIPG 2008\/}, pages 154--159.

\bibitem[Langfelder and Horvath(2007)Langfelder and
  Horvath]{langfelder07eigengene}
Langfelder, P. and Horvath, S. (2007).
\newblock {Eigengene networks for studying the relationships between
  co-expression modules}.
\newblock {\em BMC Syst. Biol.}, {\bf 1}(1), 1--54.

\bibitem[Marbach {\em et~al.}(2010)Marbach, Prill, Schaffter, Mattiussi,
  Floreano, and Stolovitzky]{marbach10revealing}
Marbach, D., Prill, R., Schaffter, T., Mattiussi, C., Floreano, D., and
  Stolovitzky, G. (2010).
\newblock {Revealing strenghts and weaknesses of methods for gene network
  inference}.
\newblock {\em PNAS\/}, {\bf 107}(14), 6286--6291.

\bibitem[Margolin {\em et~al.}(2006)Margolin, Nemenman, Basso, Wiggins,
  Stolovitzky, Dalla-Favera, and Califano]{margolin06aracne}
Margolin, A., Nemenman, I., Basso, K., Wiggins, C., Stolovitzky, G.,
  Dalla-Favera, R., and Califano, A. (2006).
\newblock {ARACNE}: an algorithm for the reconstruction of gene regulatory
  networks in a mammalian cellular context.
\newblock {\em BMC Bioinform.}, {\bf 7}(7), S7.

\bibitem[Markowetz and Spang(2007)Markowetz and Spang]{markowetz07inferring}
Markowetz, F. and Spang, R. (2007).
\newblock Inferring cellular networks - a review.
\newblock {\em BMC Bioinform.}, {\bf 8(S6)}, S5.

\bibitem[Marr {\em et~al.}(2010)Marr, Theis, Liebovitch, and
  H{\"u}tt]{marr10patterns}
Marr, C., Theis, F., Liebovitch, L., and H{\"u}tt, M.-T. (2010).
\newblock {Patterns of Subnet Usage Reveal Distinct Scales of Regulation in the
  Transcriptional Regulatory Network of \emph{Escherichia coli}}.
\newblock {\em PLoS Comput. Biol.}, {\bf 6}(7), e1000836.

\bibitem[Matthews(1975)Matthews]{matthews75comparison}
Matthews, B. (1975).
\newblock {Comparison of the predicted and observed secondary structure of T4
  phage lysozyme}.
\newblock {\em Biochim. Biophys. Acta\/}, {\bf 405}(2), 442--451.

\bibitem[Meyer {\em et~al.}(2008)Meyer, Lafitte, and Bontempi]{meyer08minet}
Meyer, P., Lafitte, F., and Bontempi, G. (2008).
\newblock {minet: A R/Bioconductor Package for Inferring Large Transcriptional
  Networks Using Mutual Information}.
\newblock {\em BMC Bioinform.}, {\bf 9}(1), 461.

\bibitem[Nemenman {\em et~al.}(2007)Nemenman, Escola, Hlavacek, Unkefer,
  Unkefer, and Wall]{nemenman07reconstruction}
Nemenman, I., Escola, G., Hlavacek, W., Unkefer, P., Unkefer, C., and Wall, M.
  (2007).
\newblock {Reconstruction of Metabolic Networks from High-Throughput Metabolite
  Profiling Data}.
\newblock {\em Ann. N.Y. Acad. Sci.}, {\bf 1115}, 102--115.

\bibitem[Peregrin-Alvarez {\em et~al.}(2009)Peregrin-Alvarez, Xiong, Su, and
  Parkinson]{alvarez09modular}
Peregrin-Alvarez, J., Xiong, X., Su, C., and Parkinson, J. (2009).
\newblock {The Modular Organization of Protein Interactions in
  \emph{Escherichia coli}}.
\newblock {\em PLOS Comput. Biol.}, {\bf 5}(10), e1000523.

\bibitem[Scanzio {\em et~al.}(2010)Scanzio, Cumani, Gemello, Mana, and
  Laface]{scanzio10parallel}
Scanzio, S., Cumani, S., Gemello, R., Mana, F., and Laface, P. (2010).
\newblock {Parallel implementation of Artificial Neural Network training for
  speech recognition}.
\newblock {\em Pattern Recogn. Lett.}, {\bf 31}(11), 1302--1309.

\bibitem[Song {\em et~al.}(2009)Song, Kolar, and Xing]{song09keller}
Song, L., Kolar, M., and Xing, E. (2009).
\newblock {KELLER: estimating time-varying interactions between genes}.
\newblock {\em Bioinformatics\/}, {\bf 25}(12), i128--i136.

\bibitem[Specht(1993)Specht]{specht93general}
Specht, D. (1993).
\newblock The general regression neural network-rediscovered.
\newblock {\em Neural Networks\/}, {\bf 6}(7), 1033--1034.

\bibitem[Steinhoff and Vingron(2006)Steinhoff and
  Vingron]{steinhoff06normalization}
Steinhoff, C. and Vingron, M. (2006).
\newblock {Normalization and quantification of differential expression in gene
  expression microarrays}.
\newblock {\em Brief. Bioinform.}, {\bf 7}(2), 166--177.

\bibitem[Stokic {\em et~al.}(2009)Stokic, Hanel, and Thurner]{stokic09fast}
Stokic, D., Hanel, R., and Thurner, S. (2009).
\newblock A fast and efficient gene-network reconstruction method from multiple
  over-expression experiments.
\newblock {\em BMC Bioinformatics\/}, {\bf 10}(1), 253.

\bibitem[Supper {\em et~al.}(2007)Supper, Spieth, and
  Zell]{supper07reconstructing}
Supper, J., Spieth, C., and Zell, A. (2007).
\newblock Reconstructing linear gene regulatory networks.
\newblock In {\em Proc. of EvoBIO2007\/}, pages 270--279. Springer-Verlag.

\bibitem[Tuna and Niranjan(2009)Tuna and Niranjan]{tuna09cross}
Tuna, S. and Niranjan, M. (2009).
\newblock {Cross-Platform Analysis with Binarized Gene Expression Data}.
\newblock In {\em PRIB\/}, pages 439--449.

\end{thebibliography}
